\documentclass[%
reprint,
superscriptaddress,
prc,
]{revtex4-2}

\usepackage{graphicx}
\usepackage{xcolor}
\usepackage{dcolumn}% Align table columns on decimal point
\usepackage[T1]{fontenc}% else broken \k in references
\usepackage{tabularx}% table options
\usepackage{mathtools}
\usepackage{amsmath,amssymb}
\usepackage{physics}
\usepackage{bm}
\usepackage{url}
\usepackage[colorlinks,hyperindex]{hyperref}
\hypersetup
{
   colorlinks,%
   citecolor=blue,%
   linkcolor=blue,%
   urlcolor=blue,%
}
\usepackage[capitalise]{cleveref}%must come after hyperref
\usepackage{lipsum}
\usepackage{blindtext}

\begin{document}

%-------------------------%
% frontmatter
\title{A Two-Component Exciton Model for $\beta$-Delayed Neutron Emission}

\author{Nico Braukman}
\affiliation{Department of Physics and Astronomy, University of Tennessee,
					Knoxville, Tennessee 37996, USA}
\affiliation{Theoretical Division, Los Alamos National Laboratory,
					Los Alamos, New Mexico 87545, USA}

\author{Toshihiko Kawano}
\affiliation{Theoretical Division, Los Alamos National Laboratory,
					Los Alamos, New Mexico 87545, USA}

\author{Robert Grzywacz}
\affiliation{Department of Physics and Astronomy, University of Tennessee,
					Knoxville, Tennessee 37996, USA}

\author{Zhengyu Xu}
\affiliation{Department of Physics and Astronomy, University of Tennessee,
					Knoxville, Tennessee 37996, USA}

\date{\today}

\begin{abstract}
A model for $\beta$-delayed neutron emission is presented in which the emission
of neutrons is allowed to occur before states populated in $\beta^-$ decay have
reached equilibrium in the so-called ``compound nucleus'' picture. The
pre-equilibrium neutron spectra are computed using a two-component exciton
model, where neutron and proton, particle and hole degrees of freedom are
treated independently. This model is implemented in the code \textsc{BeoH} to
supplement the existing Hauser-Feshbach model for statistical decay of
compound-nucleus states. The effect of a pre-equilibrium component in the total
neutron spectrum on the single- and multi-neutron emission probabilities
$P_{xn}$ is investigated for $\beta$-decay precursor nuclei in the regions
$47\leq Z\leq 49,~83\leq N\leq 86$ and $54\leq Z\leq 58,~N=125,126$. An
absolute change in the $P_{xn}$ values of up to 5-10\% is observed in the
$N=126$ region when including pre-equilibrium emission compared to the values
predicted by the Hauser-Feshbach model alone.
\end{abstract}

\maketitle
%-------------------------%

%=============================================================================%

%-------------------------%
\section{Introduction\label{sec:intro}}
%-------------------------%

\begin{figure*}
\includegraphics[width=0.85\textwidth]{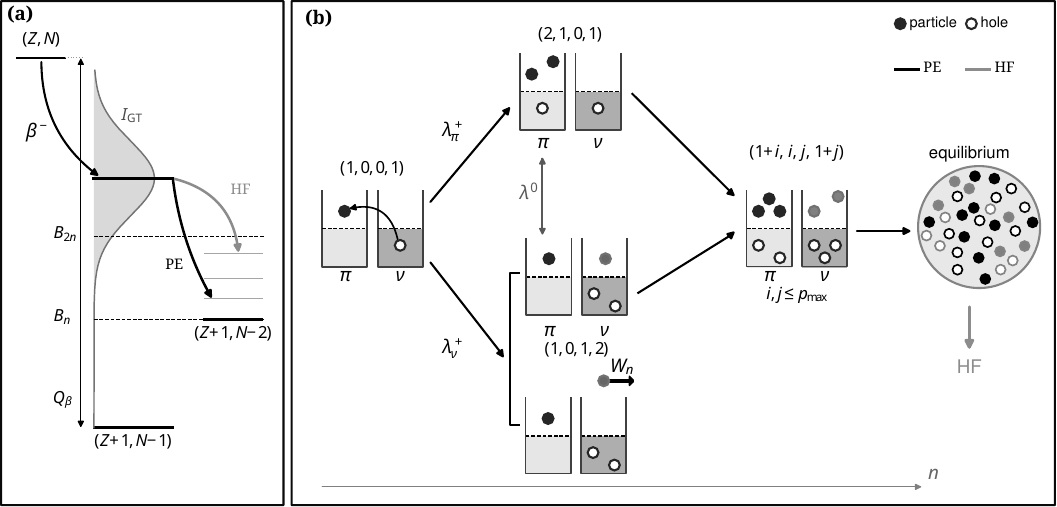}
\caption{(a) Schematic $\beta$-$xn$ decay in the ``three-step'' pre-equilibrium
(PE) plus Hauser-Feshbach (HF) model. (b) Exciton model evolution from an
initial state populated in $\beta^-$ decay. Neutron emission from equilibrated
states is treated in the HF formalism.}
\label{fig:schematic}
\end{figure*}

In neutron-rich regions of the nuclear chart, it is possible for $\beta^-$
decay to populate excited daughter states that lie above the neutron
binding energy $B_{n}$. Daughter nuclei in these states can subsequently decay
by emission of the unbound neutron \cite{Rob1939,Boh1939}, and, for
sufficiently exotic precursors, multi-neutron emission may be possible if the
$\beta^-$-decay energy window $Q_{\beta}$ is greater than the multi-neutron
binding energies $B_{xn}$, with $x=1,2,3,\ldots$ The present state of the art
in theoretical descriptions of this $\beta$-delayed (multi-)neutron decay, or
$\beta$-$xn$ decay, is to first obtain the distribution of daughter states
produced in $\beta^-$ decay using microscopic models, such as the Large-Scale
Shell Model (LSSM) or Quasiparticle Random Phase Approximation (QRPA), and
then, to determine the decay of these daughter states through statistical decay
models, as in the Hauser-Feshbach theory \cite{Hau1952}. Such ``two-step''
methods have been employed to compute the theoretical $\beta$-delayed neutron
spectra and associated observables across the nuclear chart \cite{Mum2025}.

A major assumption inherent to this method is that the states populated in
$\beta^-$ decay will reach statistical equilibrium and form a so-called
``compound nucleus'' \cite{Boh1936} before particle decay can occur. However,
from reaction experiments performed since the 1960s \cite{Hol1963,Woo1965},
it is known that, at high system energies, particle decay can occur before
equilibration. To describe this pre-equilibrium emission, Griffin introduced
the exciton model \cite{Gri1966}, in which the initial state of the system
is described by a given configuration of particles and holes (collectively
called excitons) that interact to excite more particle-hole pairs as the system
evolves towards equilibrium. At each stage of the evolution, a fraction of the
particles excited above the binding energy may be emitted from the system,
contributing to the pre-equilibrium particle spectrum. Since the work of
Griffin, further developments to the exciton model have been made, including
the separation of particle and hole \cite{Cli1971} and proton and neutron
\cite{Gup1981,Dob1983,Kal1986} degrees of freedom, and the use of modern
global optical model potentials to compute the inverse reaction cross-sections
\cite{Kon2004}.
In this work, we employ a two-component (independent protons and neutrons)
exciton model to compute the pre-equilibrium neutron spectra following
$\beta^-$ decay for the first time.
Recent experimental evidence for pre-equilibrium emission has been seen
in double-differential cross section measurements: for example, in pre-fission
neutrons from $^{239}$Pu$(n,f)$ reactions \cite{Kel2019} and in
photoneutron emission following giant-dipole-resonance excitation
\cite{Kir2020,Ngu2024,Ngu2025}. The latter is the closest analog to $\beta$
decay, since the initial $1p$-$1h$ excitation is created without an incident
particle, and all nucleons are initially bound. However, pre-equilibrium
neutron emission has not yet been experimentally demonstrated following $\beta$
decay.
As experimental facilities continue to push towards more exotic regions of
the nuclear chart \cite{Nak2017,Ma2021}, we expect to observe $\beta$-$xn$
decays of higher multiplicity and with higher-energy neutrons.
In these regions, pre-equilibrium effects on experimental observables may
be significant, and we would like to compare $\beta$-$xn$ decay measurements
with a theory that accounts for this pre-equilibrium component.

In our ``three-step'' approach to computing the total $\beta$-delayed neutron
spectra, we first obtain the population of $\beta^-$-decay daughter states
through LSSM calculations, then, we compute the pre-equilibrium spectra for
primary neutron emission, and finally, we use the Hauser-Feshbach theory to
compute the primary and multi-neutron emission spectra from the fraction of the
daughter population that survives to equilibration.
The approach is schematized in \cref{fig:schematic}.
This model was developed
for the reaction code \textsc{CoH}$_{3}$ \cite{Kaw2010,CoH}, and, in this work,
is incorporated into the derivative program \textsc{BeoH}
\cite{Oku2018,Lov2021} for $\beta$ decay, in which the Hauser-Feshbach model
for statistical
decay has already been implemented. In \cref{sec:model}, we outline the exciton
model formalism and describe its implementation in the \textsc{BeoH} program,
including a modification to the standard prescription, in which particle
emission from unbound configurations in the early stages of the evolution is
treated somewhat differently to achieve better consistency with
quantum-mechanical approaches to pre-equilibrium decay \cite{Cha1993,Sas2025}.
In \cref{sec:results}, we apply the exciton and Hauser-Feshbach models to
investigate the impact of pre-equilibrium emission on the $P_{xn}$ values and
neutron multiplicities following $\beta^-$ decay of nuclei in two regions of
the nuclear chart: (a) ``southeast'' of $^{132}$Sn, with $47\leq Z\leq 49$ and
$83\leq N\leq 86$, and (b) along the $N=125$ and 126 isotonic chains at low
proton number $54 \leq Z \leq 58$. These regions are of significant importance
to the rapid neutron capture ($r$) process for astrophysical nucleosynthesis
\cite{Bur1957,Cow2021}, where proximity to closed neutron shells at ``magic
numbers'' $N=82$ and 126 strongly affects $\beta^-$-decay rates and neutron
emission probabilities, and where, in the absence of experimental data,
accurate theoretical modeling is most crucial \cite{Mum2016}. Accurate modeling
of $\beta$-$xn$ decay in the $^{132}$Sn region is also critical for fission and
reactor applications \cite{Tob1980,Alg2010}. These regions are characterized by
high $Q_\beta$ and low $B_{xn}$ values; with the possibility of populating
states at high excitation energies, we can expect a non-trivial impact of
pre-equilibrium neutron emission in these regions. We conclude in
\cref{sec:conclusions} with some remarks on the results and enumerate possible
future improvements to the model.

%=============================================================================%

%-------------------------%
\section{Model Overview\label{sec:model}}
%-------------------------%

The two-component exciton model \cite{Kaw2010} is adapted for and implemented
in \textsc{BeoH} \cite{Oku2018,Lov2021} for $\beta$-delayed particle decay. The
model takes as input the Gamow-Teller $\beta$-decay intensity distribution
$I_\text{GT}$ as a function of excitation energy $E_\text{tot} \in (0,Q_\beta)$
in the $\beta$-decay daughter. We neglect first- and higher-order-forbidden
$\beta$ transitions and assume only allowed Gamow-Teller transitions
from a precursor state of spin $J_0$ and parity $\pi_0$, so that the
spin-parities of the $\beta$-decay daughter are $J = J_0, |J_0\pm 1|$
(no $0\to0$ transitions) and $\pi=\pi_0$. States in the $\beta$-decay
daughter within the energy bin $(E_\text{tot},E_\text{tot}+dE_\text{tot})$ are
assigned a population $d\sigma(E_\text{tot},J^\pi)$, in fraction of $\beta$
decays, according to $I_\text{GT}$, with the strength divided evenly among the
possible spins $J$.

We iterate over the states in the $\beta$-decay daughter and, for
each state, perform an exciton model calculation to compute the
pre-equilibrium particle emission spectrum. Hereafter, we will use the notation
$(p_\pi,h_\pi,p_\nu,h_\nu)$ to describe an exciton configuration consisting of
$p_{\pi(\nu)}$ particles in the proton (neutron) shell and $h_{\pi(\nu)}$ holes
in the proton (neutron) shell, relative to the precursor state. The initial
exciton configuration for the particle-emitting states populated in $\beta^-$
decay is assumed to be $(1,0,0,1)$, that is, one proton particle and one
neutron hole relative to the precursor.

\subsection{Particle Emission Rates}

The emission rate for particles of species $k$ (with proton number $Z_k$ and
neutron number $N_k$) and energy $E_k$ from a state at energy $E_\text{tot}$
with a given exciton configuration is given by the following formula
\cite{Kal1986}, where $s_k$ is the intrinsic spin of the ejectile, $\mu_k$ the
reduced mass of the ejectile-plus-residue system,
$\sigma_{\text{inv}}^{(R)}(E_k)$ the cross section for the inverse reaction (a
projectile $k$ incident on a target equivalent to the residue nucleus in a
state at energy $E_R=E_\text{tot}-B_k-E_k$ with spin-parity $J_R^{\pi_R}$,
where $B_k$ is the binding energy of $k$ particles), and $\omega$ is the
particle-hole state density for a given energy and exciton configuration,
\begin{align}\label{eq:EmissionRate}
W_k(p_\pi,h_\pi,p_\nu,h_\nu;E_\text{tot}) =
\frac{2s_k+1}{\pi^2\hbar^3} \mu_k E_k \sigma_\text{inv}^{(R)}(E_k)\nonumber\\
\times\frac{\omega(p_\pi-Z_k,k_\pi,p_\nu-N_k,h_\nu;E_R)
	}{\omega(p_\pi,h_\pi,p_\nu,h_\nu;E_\text{tot})}.
\end{align}
The total emission rate $W(p_\pi,h_\pi,p_\nu,h_\nu;E_\text{tot})$ is the sum
and integral of $W_k$ over all possible ejectile species $k$ and energies
$E_k$.

In practice, we compute the inverse cross section
$\sigma_\text{inv}^{(R)}(E_k)$ using transmission coefficients $T_{\ell j}$
from the Koning-Delaroche optical model potential \cite{Kon2003}. We compute
\begin{align}\label{eq:InverseXsec}
\sigma_\text{inv}^{(R)}(E_k) =
\frac{\pi}{k_R^2}
\sum_{\ell j} \frac{2j+1}{2s_k+1} T_{\ell j}^{(R)},
\end{align}
the cross section for incident particles of energy $E_k$, wavenumber $k_R$, and
orbital and total angular momenta $\ell$ and $j$, incident on a target nucleus
in the state $E_R, J_R^{\pi_R}$, to form a compound system in the state 
$E_\text{tot},J^\pi$.

We compute the particle-hole state densities at energy $E$ and given exciton
configuration using the following formula \cite{Kal1986,Kon2004},
\begin{align}
\omega(p_\pi,h_\pi,p_\nu,h_\nu;E) &=
\frac{g_\pi^{n_\pi} g_\nu^{n_\nu}}{p_\pi!h_\pi!p_\nu!h_\nu!(n-1)!}\nonumber\\
&\times \left( U - A(p_\pi,h_\pi,p_\nu,h_\nu) \right)^{n-1}\nonumber\\
&\times f(n,h,U,V),
\end{align}
where $p=p_\pi+p_\nu$ is the total number of particles, $h=h_\pi+h_\nu$ the
total number of holes, $n_\pi=p_\pi+h_\pi$ the number of proton excitons, 
$n_\nu = p_\nu + h_\nu$ the number of neutron excitons, and $n=p+h=n_\pi+n_\nu$
the total exciton number. The single-particle state densities are given by
\cite{Kaw2015}
\begin{subequations}
\begin{align}
g_\pi &= Z/(15.7~\text{MeV})\\
g_\nu &= N/(19.2~\text{MeV}).
\end{align}
\end{subequations}

The pairing-corrected energy is $U \equiv E - P_{p,h}$, where we take the
pairing correction according to \cite{Fu1984},
\begin{subequations}
\begin{align}
P_{p,h} &= \begin{cases}
\Delta\left[
1 - \left( 0.996 - 1.76(\frac{n}{n_c})^{1.6}
	(\frac{E_\text{CN}}{\Delta})^{-0.68} \right)^2
\right],\\
\qquad \frac{E_\text{CN}}{\Delta} \geq 0.716 + 2.44(\frac{n}{n_c})^{2.17}\\
\Delta,\quad\text{otherwise}
\end{cases}\\
\Delta &= \chi \frac{12}{\sqrt{A}}~\text{MeV},\\
n_c &= 1.386 (g_\pi + g_\nu) \sqrt{\frac{\Delta}{g_\pi+g_\nu}}~\frac{8}{7},
\end{align}
\end{subequations}
where $\chi = 0,1,2$ for odd-odd, odd-mass, and even-even nuclei,
respectively.

The Pauli correction term $A(p_\pi,h_\pi,p_\nu,h_\nu)$ accounts for the minimum
energy required, under the Pauli exclusion principle, to form a given exciton
configuration. We take the expression from \cite{Kal1986},
\begin{align}\label{eq:Pauli}
A(p_\pi,h_\pi,p_\nu,h_\nu) =
	\frac{\text{max}(p_\pi,h_\pi)^2}{g_\pi} -
	\frac{p_\pi^2 + h_\pi^2 + n_\pi}{4 g_\pi}\nonumber\\
	+ \frac{\text{max}(p_\nu,h_\nu)^2}{g_\nu} -
	\frac{p_\nu^2 + h_\nu^2 + n_\nu}{4 g_\nu}.
\end{align}

Finally, the finite-well correction term $f(n,h,U,V)$ accounts for the fact
that the holes are bounded by the bottom of a finite potential well, whose
depth is equal to the Fermi energy $V = E_F = 38.0$ MeV. The correction term is
given by \cite{Kal1986}
\begin{align}\label{eq:FiniteWell}
f(n,h,U,V) =
1 + \sum_{i=1}^h (-1)^i {{h}\choose{i}}
\left(1-\frac{iV}{U}\right)^{n-1} \nonumber\\ \times \Theta(U-iV),
\end{align}
where $\Theta$ is the Heaviside step function. In the case of reactions due to
an incident projectile, Kalbach \cite{Kal1985,Kal2000} introduces an effective
well depth for the $h=1$ case, due to the fact that, at the initial stages of
the reaction, the incident particle sees a shallower potential at the surface
of the target nucleus. In the $\beta$-decay case, however, the inital neutron
hole (and any subsequent holes) can be found at any depth in the potential
well, so we forgo this surface correction and take $V = E_F$ for any $h$.

\subsection{Internal Transition Rates}

The flux passing through a given exciton configuration that does not result in
pre-equilibrium particle emission instead results in transition to a different
exciton configuration. These internal transitions are governed by a two-body
residual interaction \cite{Gri1966} which can change the proton or neutron
exciton number by $0$ or $\pm 2$ through particle-hole pair production,
annihilation, and charge exchange (conversion of a proton or neutron
particle-hole pair into a pair of the other species). In this work, we make the
assumption that, in the early stages of the exciton model evolution, the
probability of transition to a configuration with fewer excitons is negligible,
since there are many more configurations available at higher exciton number
(known as the ``never come back'' approximation \cite{Kon2004}). For similar
reasons, we further assume that at most one charge-exchange interaction can
occur for a given total exciton number $n$ before pair creation takes the
system to an $n+2$ configuration.

The matrix elements for the two-body interaction that governs the internal
transitions are given by
\begin{subequations}
\begin{align}
M_{\pi\pi}^2 &= M_{\nu\nu}^2 = M^2\\
M_{\pi\nu}^2 &= M_{\nu\pi}^2 = R_{\pi\nu}M^2
\end{align}
\end{subequations}
where \cite{Kon2004}
\begin{align}
M^2 = \frac{1}{A^3}\left(
	6.8~\text{MeV}^2 + 
	\frac{
		4.2\times10^5~\text{MeV}^5
	}{
		\left( \frac{E_\text{tot}}{n} + 10.7~\text{MeV} \right)^3
	}
\right)
\end{align}
is a phenomenological average matrix element for a nucleus of $A$ nucleons in a
state at energy $E_\text{tot}$ with $n$ total excitons, and $R_{\pi\nu}=1.5$
\cite{Kon2004} is the ratio of strengths of proton-neutron to like-nucleon
interactions.

The transition rates $\lambda_\pi^+,\lambda_\nu^+,\lambda_\pi^0,\lambda_\nu^0$
for proton particle-hole pair creation, neutron particle-hole pair creation,
conversion of a neutron particle-hole pair into a proton particle-hole pair,
and conversion of a proton particle-hole pair into a neutron particle-hole
pair, respectively, are given by the following formulas \cite{Kon2004}. For
neutron particle-hole pair creation, we have
\begin{widetext}
\begin{align}
\lambda_\nu^+(p_\pi,h_\pi,p_\nu,h_\nu) &=
\frac{1}{\omega(p_\pi,h_\pi,p_\nu,h_\nu;E_\text{tot})} \times\nonumber\\
&\Bigg\{
	\int_{L_1^{p_\nu}}^{L_2^{p_\nu}} du \left[
		\lambda_{\nu\nu}^{1p}(u)
		~\omega(p_\pi,h_\pi,p_\nu-1,h_\nu;E_\text{tot}-u)
		~\omega(0,0,1,0;u)
	\right]\nonumber\\
	&+
	\int_{L_1^{h_\nu}}^{L_2^{h_\nu}} du \left[
		\lambda_{\nu\nu}^{1h}(u)
		~\omega(p_\pi,h_\pi,p_\nu,h_\nu-1;E_\text{tot}-u)
		~\omega(0,0,0,1;u)
	\right]\nonumber\\
	&+
	\int_{L_1^{p_\pi}}^{L_2^{p_\pi}} du \left[
		\lambda_{\pi\nu}^{1p}(u)
		~\omega(p_\pi-1,h_\pi,p_\nu,h_\nu;E_\text{tot}-u)
		~\omega(1,0,0,0;u)
	\right]\nonumber\\
	&+
	\int_{L_1^{h_\pi}}^{L_2^{h_\pi}} du \left[
		\lambda_{\pi\nu}^{1h}(u)
		~\omega(p_\pi,h_\pi-1,p_\nu,h_\nu;E_\text{tot}-u)
		~\omega(0,1,0,0;u)
	\right]
\Bigg\}.
\end{align}
\end{widetext}

That is, the neutron particle-hole pair production rate is the sum of the
production rates due to neutron interactions with a neutron particle, neutron
hole, proton particle, and proton hole in the current configuration of the
state. The integrations run over the energy $u$ available to an individual
particle or hole degree of freedom. The analogous expression for
$\lambda_\pi^+$ is obtained by interchanging neutron and proton variables in
the above. The limits of integration are given below, where
$A(p_\pi,h_\pi,p_\nu,h_\nu)$ is the Pauli correction term \cref{eq:Pauli}.

\begin{subequations}
\begin{align}
\begin{cases}
L_1^{p_\nu} = A(p_\pi,h_\pi,p_\nu+1,h_\nu+1) - A(p_\pi,h_\pi,p_\nu-1,h_\nu)\\
L_2^{p_\nu} = E_\text{tot}- A(p_\pi,h_\pi,p_\nu-1,h_\nu)
\end{cases}\\
\begin{cases}
L_1^{h_\nu} = A(p_\pi,h_\pi,p_\nu+1,h_\nu+1) - A(p_\pi,h_\pi,p_\nu,h_\nu-1)\\
L_2^{h_\nu} = E_\text{tot}- A(p_\pi,h_\pi,p_\nu,h_\nu-1)
\end{cases}\\
\begin{cases}
L_1^{p_\pi} = A(p_\pi+1,h_\pi+1,p_\nu,h_\nu) - A(p_\pi-1,h_\pi,p_\nu,h_\nu)\\
L_2^{p_\pi} = E_\text{tot}- A(p_\pi-1,h_\pi,p_\nu,h_\nu)
\end{cases}\\
\begin{cases}
L_1^{h_\pi} = A(p_\pi+1,h_\pi+1,p_\nu,h_\nu) - A(p_\pi,h_\pi-1,p_\nu,h_\nu)\\
L_2^{h_\pi} = E_\text{tot}- A(p_\pi,h_\pi-1,p_\nu,h_\nu)
\end{cases}
\end{align}
\end{subequations}

The factors $\lambda_{\nu\nu}^{1p}(u),\lambda_{\nu\nu}^{1h}(u),$ etc., are the
collision probability rates given by Fermi's golden rule,

\begin{subequations}
\begin{align}
\lambda_{\nu\nu}^{1p}(u)&=\frac{2\pi}{\hbar}M_{\nu\nu}^2\omega(0,0,2,1;u),\\
\lambda_{\nu\nu}^{1h}(u)&=\frac{2\pi}{\hbar}M_{\nu\nu}^2\omega(0,0,1,2;u),\\
\lambda_{\pi\nu}^{1p}(u)&=\frac{2\pi}{\hbar}M_{\pi\nu}^2\omega(1,0,1,1;u),\\
\lambda_{\pi\nu}^{1h}(u)&=\frac{2\pi}{\hbar}M_{\pi\nu}^2\omega(0,1,1,1;u).
\end{align}
\end{subequations}

For example, $\lambda_{\nu\nu}^{1p}(u)$ is the probability rate for a neutron
particle with energy $u$ to produce a neutron particle-hole pair by interacting
with the neutron shell; $\lambda_{\pi\nu}^{1h}(u)$ is the probability rate for
a proton hole with energy $u$ to produce a neutron particle-hole pair by
interacting with the neutron shell; etc. The analogous expressions for
$\lambda_{\pi\pi}^{1p(1h)}(u),\lambda_{\nu\pi}^{1p(1h)}(u)$ which appear in the
formula for $\lambda_\pi^+$ are obtained by interchanging neutron and proton
variables in the above.

The transition rate for conversion of a neutron particle-hole pair into a
proton particle-hole pair is given by
%\begin{widetext}
\begin{align}
\lambda_\pi^0(p_\pi,&h_\pi,p_\nu,h_\nu) =
\frac{1}{\omega(p_\pi,h_\pi,p_\nu,h_\nu;E_\text{tot})} \times\nonumber\\
&\int_{L_1^{p_\nu}}^{L_2^{p_\nu}} du \big[
\lambda_{\nu\pi}^{1p1h}(u)\nonumber\\
&\qquad\qquad\times\omega(p_\pi,h_\pi,p_\nu-1,h_\nu-1;E_\text{tot}-u)
\nonumber\\
&\qquad\qquad\times\omega(0,0,1,1;u)
\big].
\end{align}
%\end{widetext}
The limits of integration are
\begin{align}
\begin{cases}
L_1^{p_\nu} = A(p_\pi,h_\pi,p_\nu,h_\nu) - A(p_\pi,h_\pi,p_\nu-1,h_\nu-1)\\
L_2^{p_\nu} = E_\text{tot} - A(p_\pi,h_\pi,p_\nu-1,h_\nu-1),
\end{cases}
\end{align}
and the collision probability rate for a proton-neutron interaction to result
in the annihilation of a neutron particle-hole pair and production of a proton
particle-hole pair is
\begin{align}
\lambda_{\nu\pi}^{1p1h}(u)=\frac{2\pi}{\hbar}M_{\nu\pi}^2\omega(1,1,0,0;u).
\end{align}

The analogous expression for the proton-to-neutron conversion rate
$\lambda_\nu^0$ is obtained by interchanging proton and neutron variables in
the above. The lifetime $\tau$ of a state at energy $E_\text{tot}$ with a
given exciton configuration is the inverse of the sum of all the internal
transition rates and particle emission rates from this state:
\begin{align}
\tau(p_\pi,h_\pi,p_\nu,h_\nu;E_\text{tot}) = 1/\big(
\lambda_\pi^+(p_\pi,h_\pi,p_\nu,h_\nu;E_\text{tot})\nonumber\\
+\lambda_\nu^+(p_\pi,h_\pi,p_\nu,h_\nu;E_\text{tot})\nonumber\\
+\lambda_\pi^0(p_\pi,h_\pi,p_\nu,h_\nu;E_\text{tot})\nonumber\\
+\lambda_\nu^0(p_\pi,h_\pi,p_\nu,h_\nu;E_\text{tot})\nonumber\\
+W(p_\pi,h_\pi,p_\nu,h_\nu;E_\text{tot})
\big)
\end{align}

\subsection{Emitted Particle Spectra}

The exciton model calculation proceeds as follows from the initial exciton
configuration $(1,0,0,1)$ of a state $E_\text{tot}, J^\pi$ populated in
$\beta^-$ decay. First, the internal transition rates $\lambda_{\pi(\nu)}^+,
\lambda_{\pi(\nu)}^0$ and emission rates $W_k$ are calculated for the initial
and all subsequent exciton configurations of the state: we iterate
over the configurations $(1+i,0+i,0+j,1+j)$, where $i,j=0,1,\ldots,
p_\text{max}$, and where $p_\text{max}\equiv 6$ is a truncation of the
particle-hole pairs that can be created. Beyond $p_\text{max}$, the system is
assumed to have equilibrated, and the particle emission is handled in
\textsc{BeoH} by the Hauser-Feshbach model for statistical decay.

Next, we compute the occupation $\eta_{ij}$ of each configuration, accounting
for the flux due to particle emission and internal transitions. We set
$\eta_{00} \equiv 1$, and compute the higher-exciton occupations recursively
\cite{Kon2004}:
\begin{align}\label{eq:Pop}
\eta_{ij} &=
  \eta_{i-1,j}\lambda^+_{\pi(i-1,j)}\tau_{i-1,j}\nonumber\\
&+\eta_{i,j-1}\lambda^+_{\nu(i,j-1)}\tau_{i,j-1}\nonumber\\
&+\lambda^0_{\nu(i-1,j+1)}\tau_{i-1,j+1}\bigg(
	\eta_{i-2,j+1}\lambda^+_{\pi(i-2,j+1)}\tau^\prime_{i-2,j+1}\nonumber\\
	&\qquad\qquad\qquad\qquad\qquad
	+\eta_{i-1,j}\lambda^+_{\nu(i-1,j)}\tau^\prime_{i-1,j}
\bigg)\nonumber\\
&+\lambda^0_{\pi(i+1,j-1)}\tau_{i+1,j-1}\bigg(
	\eta_{i+1,j-2}\lambda^+_{\nu(i+1,j-2)}\tau^\prime_{i+1,j-2}\nonumber\\
	&\qquad\qquad\qquad\qquad\qquad
	+\eta_{i,j-1}\lambda^+_{\pi(i,j-1)}\tau^\prime_{i,j-1}
\bigg),
\end{align}
where we write $\lambda_{\alpha(i,j)}^{x} \equiv \lambda_\alpha^x(1+i,0+i,0+j,
1+j;E_\text{tot})$ and $\tau_{ij} \equiv \tau(1+i,0+i,0+j,1+j;E_\text{tot})$
for ease of notation. We define the lifetime neglecting charge exchange to be
\begin{align}
\tau^\prime(p_\pi,h_\pi,p_\nu,h_\nu;E_\text{tot}) = 1/\big(
\lambda_\pi^+(p_\pi,h_\pi,p_\nu,h_\nu;E_\text{tot})\nonumber\\
+\lambda_\nu^+(p_\pi,h_\pi,p_\nu,h_\nu;E_\text{tot})\nonumber\\
+W(p_\pi,h_\pi,p_\nu,h_\nu;E_\text{tot})
\big),
\end{align}
thus accounting for the approximation that only one charge-exchange interaction
can occur for each total exciton number $n$. In each term of \cref{eq:Pop}, the
indices specifying the exciton configuration are assumed to be in the range
$[0,p_\text{max}]$, otherwise, the term is zero. Since the occupation of the
$i,j$ configuration depends on the occupation at $i-2$ and $j-2$,
we re-computes \cref{eq:Pop} until convergence.

Finally, the exciton occupations must be converted into the appropriate
particle spectra. In this step, again, we iterate over each
configuration $i,j$, and calculate the particle spectra $d\sigma_k/dE_k$
for each possible ejectile species $k$ with energy $E_k$ as
\begin{align}\label{eq:Spectra}
\frac{d\sigma_k}{dE_k}\bigg|_{E_\text{tot},J^\pi}
= \frac{d\sigma}{dE_\text{tot}}dE_\text{tot} ~
\sum_{i,j=0}^{p_\text{max}} W_{k(ij)} ~ \tau_{ij} ~ \eta_{ij}
\end{align}
where $W_{k(ij)} = W_k(1+i,0+i,0+j,1+j;E_k)$ and $d\sigma$ is
the population (in fraction of $\beta$ decays) of states at energy within
$dE_\text{tot}$ of $E_\text{tot}$ and spin-parity $J^\pi$ from which particles
are emitted. The units of $d\sigma_k/dE_k$ are fraction of $\beta$ decays per
unit ejectile energy.

In \textsc{BeoH}, the particle spectra \cref{eq:Spectra} are
calculated for every inital state $E_\text{tot},J^\pi$. We store a
cumulative spectrum
\begin{align}\label{eq:SumSpec}
\frac{d\sigma_k}{dE_k} =
\sum_{J=|J_0-1|}^{J_0+1}
\int_0^{Q_\beta} dE_\text{tot} \bigg[
\frac{d\sigma}{dE_\text{tot}}
\sum_{i,j=0}^{p_\text{max}}
W_{k(ij)} ~\tau_{ij}~\eta_{ij}
\bigg],
\end{align}
which is the sum of particle spectra from all states populated in $\beta$
decay, weighted by their respective intensities. A trivial sum over the
parity-conserved states, $\sum_{\pi}\delta_{\pi\pi_0}$, has been omitted for
simplicity.

The total pre-equilibrium emission probability from a state
$E_\text{tot},J^\pi$ is the sum of the spectra for all ejectiles $k$,
integrated over the ejectile energy $E_k$,
\begin{align}\label{eq:PEtotPerE}
d\sigma_\text{PE}(E_\text{tot},J^\pi) = 
\sum_k \int_0^{E_\text{tot}-B_k} dE_k
\frac{d\sigma_k}{dE_k}\bigg|_{E_\text{tot},J^\pi}.
\end{align}
The total pre-equilibrium emission probability from all states populated
in $\beta$ decay is obtained by integrating \cref{eq:PEtotPerE} over
$E_\text{tot}$ and summing over the allowed $J$ values,
\begin{align}\label{eq:PEtot}
\sigma_\text{PE} &= \sum_{J=|J_0-1|}^{J_0+1} \int_0^{Q_\beta} dE_\text{tot}
\bigg\{
	\frac{d\sigma}{dE_\text{tot}}\nonumber\\
	&\times \sum_k \int_0^{E_\text{tot}-B_k} dE_k
		\sum_{i,j=0}^{p_\text{max}} W_{k(ij)}~\tau_{ij}~\eta_{ij}
\bigg\}.
\end{align}

If the Hauser-Feshbach statistical particle-emission spectra are subsequently
calculated by \textsc{BeoH}, the input population of the $\beta$-decay
daughter states is scaled by $(1-\sigma_\text{PE})$ to account for the fraction
of the population that has already decayed by pre-equilibrium emission.

\subsection{Treatment of Discrete Levels\label{subsec:discrete}}

The preceeding formulations assume that the energies $E_R$ of the $k$-particle
emission residue are binned continuously in the range $(0,Q_\beta-B_k)$. If the
first $N$ low-lying levels $\mathcal{E}_R^1,\ldots,\mathcal{E}_R^N$ (with
spins $J_R^1,\ldots,J_R^N$ and parities $\pi_R^1,\ldots\pi_R^N$) in the
residue are known, then the particle emission spectra populating these levels
are calculated as follows. For each particle-emitting state $E_\text{tot},
J^\pi$, the continuum spectrum \cref{eq:Spectra} in the ejectile energy range
$E_k \in (E_\text{tot}-B_k-\mathcal{E}_R^N, E_\text{tot}-B_k)$ is distributed
among the discrete levels, weighted by the inverse reaction cross section
$\sigma_\text{inv}^{(n)}$ to each level $n=1,\ldots,N$,
\begin{subequations}
\begin{align}\label{eq:DiscreteSpectra}
d\sigma_k^{(n)}(E_\text{tot},J^\pi) &=
d\sigma_k^\text{disc} ~
\frac{\sigma_\text{inv}^{(n)}}{\sum_{n=1}^N \sigma_\text{inv}^{(n)}},
	\text{ where}\\
d\sigma_k^\text{disc} &=
\int_{E_\text{tot}-B_k-\mathcal{E}_R^N}^{E_\text{tot}-B_k}
dE_k ~ \frac{d\sigma_k}{dE_k}\bigg|_{E_\text{tot},J^\pi}
\end{align}
\end{subequations}

Then, when computing the total spectrum \cref{eq:PEtotPerE} and 
emission probability \cref{eq:PEtot}, we separate the discrete and
continuous regions,
\begin{align}
d\sigma_\text{PE}(E_\text{tot},J^\pi) = \sum_k
\sum_{n=1}^N d\sigma_k^{(n)}(E_\text{tot},J^\pi) + \nonumber\\
\sum_k \int_0^{E_\text{tot}-B_k-\mathcal{E}_R^N} dE_k
\frac{d\sigma_k}{dE_k}\bigg|_{E_\text{tot},J^\pi},
\end{align}
and
\begin{align}
\sigma_\text{PE} = \sum_{J=|J_0-1|}^{J_0+1} \int_0^{Q_\beta}
dE_\text{tot} \bigg\{
	\sum_k\sum_{n=1}^N d\sigma_k^{(n)}(E_\text{tot},J^\pi) + \nonumber\\
	\frac{d\sigma}{dE_\text{tot}}\sum_k\int_0^{E_\text{tot}-B_k-\mathcal{E}_R^N}
	dE_k \sum_{i,j=0}^{p_\text{max}} W_{k(ij)}~\tau_{ij}~\eta_{ij}
\bigg\}.
\end{align}

\subsection{Modified Emission from Unbound Configurations}

In the case of $\beta$-delayed neutron emission, we make a modification
to the exciton model evolution that enhances the neutron-emission probability
from unbound configurations in the early stages of the interaction.
Such an enhancement is suggested by quantum-mechanical approaches to
pre-equilibrium decay (as opposed to the semiclassical approach employed
in the exciton model), where, for neutron-induced reactions, it has been found
that the majority of pre-equilibrium emission occurs in the first step of the
reaction \cite{Cha1993,Sas2025}. The enhanced early emission results in a
moderately harder neutron spectrum with a stronger pre-equilibrium component
relative to the statistical component.
An enhanced pre-equilibrium component has also been observed in
photoneutron emission data \cite{Ngu2025}.
In this modified exciton model, the system evolution proceeds as follows.

In the first step of the interaction, the initial configuration $(1,0,0,1)$
cannot result in neutron emission, since there are no neutron particle degrees
of freedom. This configuration can evolve by (a) proton particle-hole pair
creation to the configuration $(2,1,0,1)$, with probability
\begin{align}
f_\pi = \frac{\lambda_{\pi(00)}^+}{
	\lambda_{\pi(00)}^+ + \lambda_{\nu(00)}^+},
\end{align}
or (b) neutron particle-hole pair creation to the configuration $(1,0,1,2)$,
with probability $f_\nu = 1-f_\pi$. In the case of neutron particle-hole
pair creation, the fraction of neutron particles in the bound energy region is
the ratio of state densities \cite{Kaw2020}
\begin{align}
r = \frac{\omega(1,0,1,2; B_n)}{\omega(1,0,1,2; E_\text{tot})},
\end{align}
and the fraction of unbound neutron particles is $1-r$.

In the modified exciton model, we treat the emission of these early unbound
neutrons differently. We assume that the internal transition rates from the
unbound configurations are hindered by factors
$\delta_{\pi(\nu)}^+,\delta_{\pi(\nu)}^0$, so that the lifetime of the unbound
$(1,0,1,2)$ configurations is
\begin{align}
\tau_{01}^\text{unb} &= 1/\left(
	\delta_\pi^+\lambda_\pi^+ + 
	\delta_\nu^+\lambda_\nu^+ + 
	\delta_\pi^0\lambda_\pi^0 + 
	\delta_\nu^0\lambda_\nu^0 + 
	W_{(01)}
\right).
\end{align}

In this work, we investigate the sensitivity of the results to a (somewhat
arbitrary) range of hindrance values $\delta_{\pi(\nu)}^+ = \delta_{\pi(\nu)}^0
\equiv \delta = 0.05,~0.10,~0.20,~0.50$. We further assume that any flux from
the unbound $(1,0,1,2)$ configurations that does not go to neutron emission
will instead evolve all the way to equilibrium, and therefore, the spectra from
subsequent configurations are not calculated. The validity of this assumption
decreases as $\delta$ approaches $1.0$. The particle emission spectrum
from the unbound $(1,0,1,2)$ configurations is computed as in
\cref{eq:Spectra}, with the above value of $\tau_{01}$, and with $\eta_{01}=1,~
\eta_{ij\neq01}=0$. The spectrum is scaled by an overall factor
$f_\nu\times(1-r)$. The neutron spectra resulting from the evolution of bound
$(1,0,1,2)$ configurations and all $(2,1,0,1)$ configurations are computed as
in the original exciton model, with overall scaling factors $f_\nu\times r$ and
$f_\pi$, respectively.

%=============================================================================%

\begin{figure}
\includegraphics[width=0.47\textwidth]{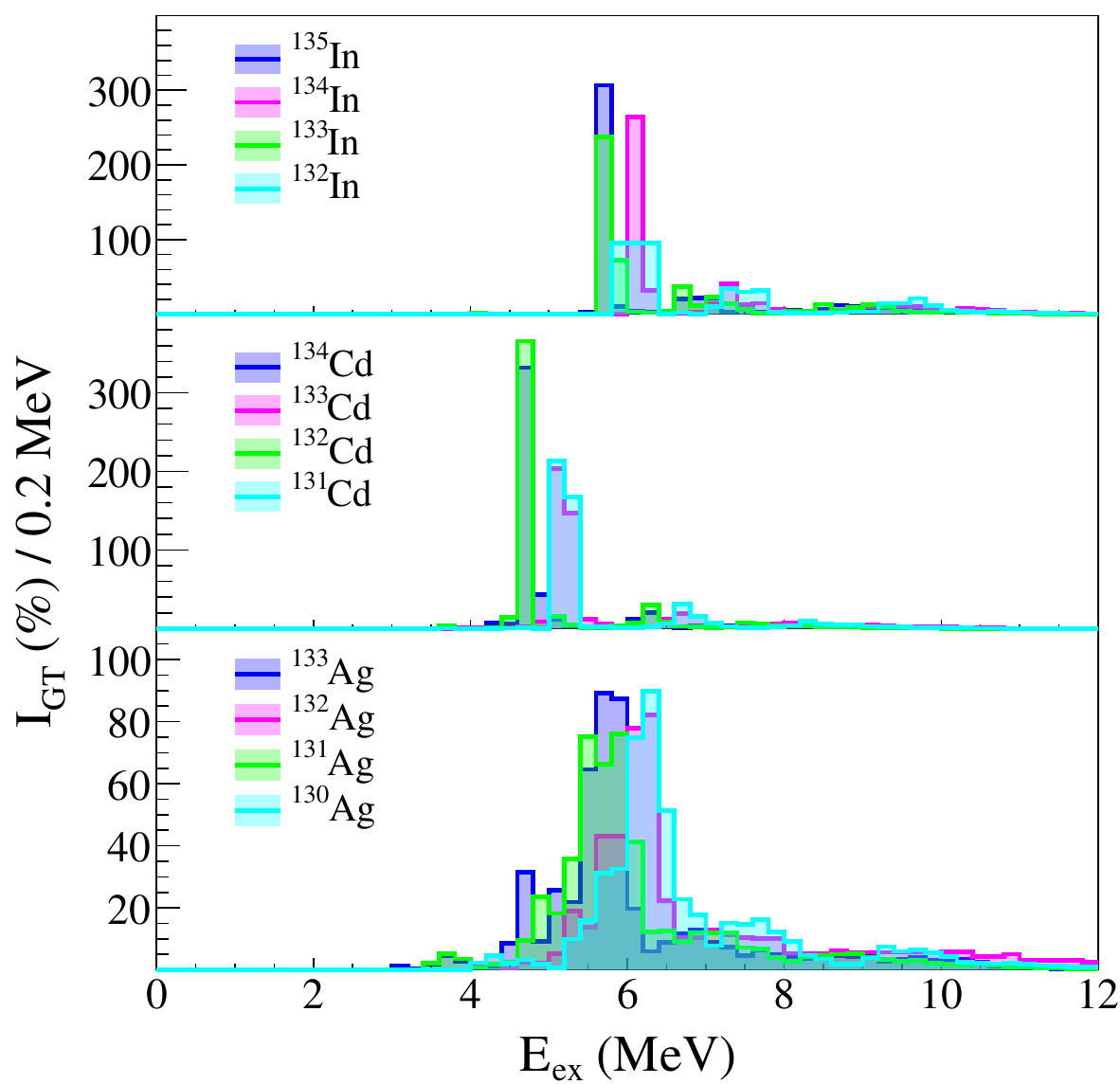}
\caption{
Gamow-Teller intensity distribution $I_\text{GT}$ for $\beta$ decay of the
indicated nuclei in the region southeast of $^{132}$Sn, from LSSM
calculations (see text for details).
}
\label{fig:IbetaSn132}
\end{figure}

\begin{figure*}[t]
\includegraphics[width=1.0\textwidth]{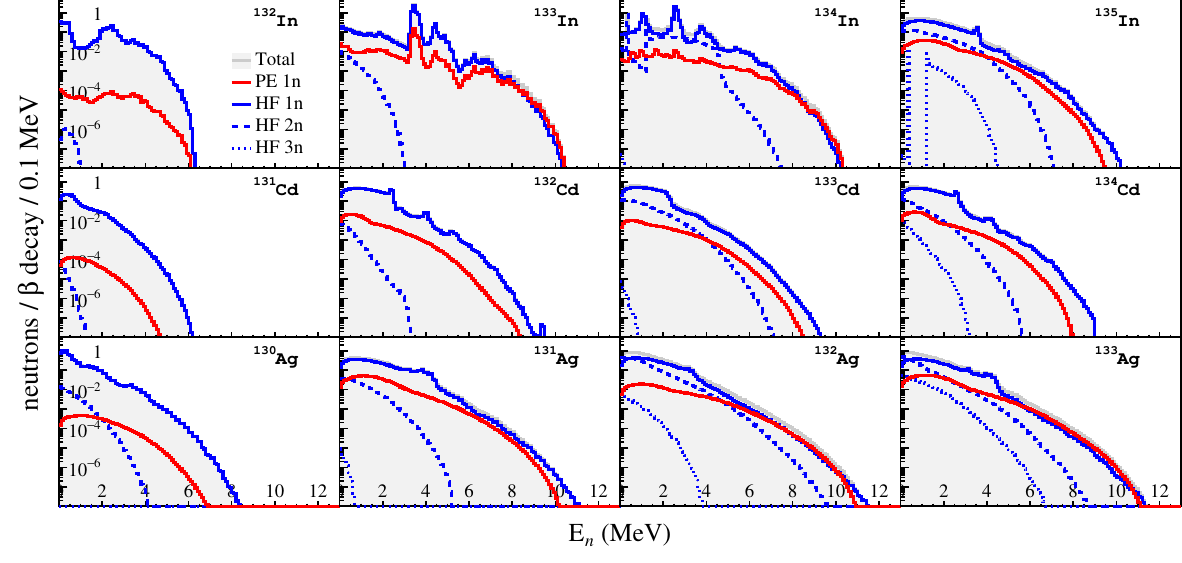}
\caption{
Pre-equilibrium (PE) and Hauser-Feshbach (HF) $\beta$-delayed neutron spectra
for nuclei southeast of $^{132}$Sn. The total spectrum (gray) is the sum of all
statistical $xn$ and pre-equilibrium $1n$ spectra. In all spectra, the
pre-equilibrium component is calculated with an internal transition hindrance
factor $\delta=0.1$.
}
\label{fig:specSn132}
\end{figure*}

%-------------------------%
\section{Results\label{sec:results}}
%-------------------------%

To illustrate the effects of pre-equilibrium emission on $\beta$-delayed
neutron spectra, we consider the $\beta$-$xn$ decay of nuclei in two
neutron-rich regions of the nuclear chart: (a) $47\leq Z \leq 49,~83\leq N\leq
86$, just ``southeast'' of doubly-magic $^{132}$Sn, and (b) $54\leq Z \leq 58,
~N=125,126$, proton-deficient isotones at and just below neutron magic number
126. We use the methods described in \cref{sec:model} to derive the
$\beta$-delayed single- and multi-neutron emission probabilities $P_{xn}$ by
integrating the appropriate spectra. We compare these probabilities to those
derived from a pure Hauser-Feshbach model for statistical decay with no
pre-equilibrium component. Although the exciton model described in
\cref{sec:model} only gives the pre-equilibrium spectrum for primary neutron
emission, the effects on the population of the $\beta$-$1n$ daughter will
propagate to the populations of the higher-$xn$ daughters, whose populations
are computed using the established Hauser-Feshbach model of \textsc{BeoH}.

\begin{figure*}
\includegraphics[width=1.0\textwidth]{%
	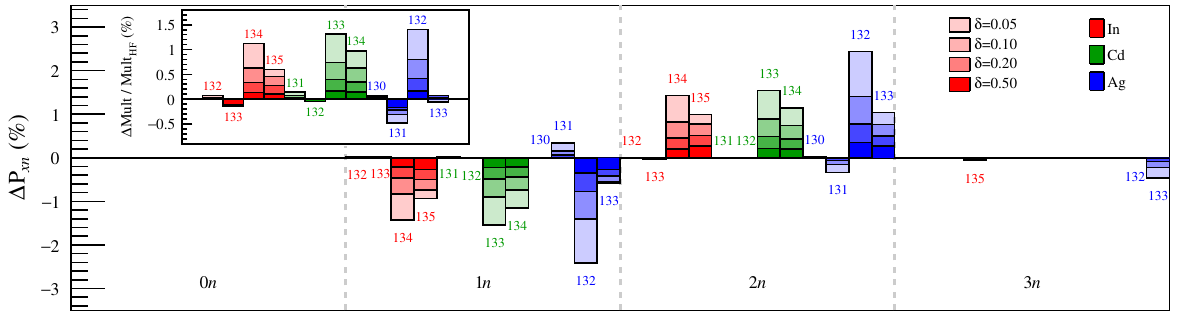}
\caption{
Difference $\Delta P_{xn}$ in predicted $xn$-emission probabilities, with and
without a pre-equilibrium component, for nuclei southeast of $^{132}$Sn.
The inset shows the difference in
neutron multiplicity relative to that predicted by pure Hauser-Feshbach decay,
$\Delta\text{Mult}/\text{Mult}_\text{HF} = (\text{Mult}_\text{HF+PE}
- \text{Mult}_\text{HF})/\text{Mult}_\text{HF}$. Bars of different
opacity correspond to calculations assuming different values of the internal
transition hindrance parameter $\delta$. The red, green, and blue colors
correspond to In, Cd, and Ag isotopes, respectively, with the bars labeled by
mass number $A$.
}
\label{fig:DPxnSn132}
\end{figure*}

\subsection{Nuclei Southeast of $^{132}$Sn}

In region southeast of doubly-magic $^{132}$Sn, we encounter high $Q_{\beta}$
values $\gtrsim 10$ MeV and low neutron binding energies
$B_{1n} \lesssim 3$ MeV; $B_{2n}\lesssim 5$ MeV \cite{Wan2021}. The possible
population of highly-excited states in the $\beta^-$-decay daughters makes this
region an ideal place to investigate pre-equilibrium effects on $\beta$-$xn$
decay.
As input to the \textsc{BeoH} exciton model, we give the theoretical
Gamow-Teller intensity distributions $I_\text{GT}$ shown in
\cref{fig:IbetaSn132} for In ($Z=49$), Cd ($Z=48$), and Ag ($Z=47$) isotopes
with $83\leq N\leq 86$. The Gamow-Teller strengths were computed in the LSSM
framework with the N$^3$LO interaction \cite{Ent2003}.
A comparison of the Gamow-Teller strength distributions $B(\text{GT})$ produced
by this interaction to experimental data on neutron-unbound states is given,
e.g., in \cite{Xu2023b,Mad2026} for $\beta$-$n$ decay of $^{133}$In and
$^{132}$Cd, respectively. Discrete levels in the daughter nuclei are taken from
the Reference Input Parameter Library (RIPL2) \cite{RIPL2} where available.

The total $\beta$-delayed neutron spectra including pre-equilibrium primary
neutron emission are shown in \cref{fig:specSn132}. For the $N=83$ isotones
(leftmost column of the figure), the pre-equilibrium component of the total
spectrum is orders of magnitude smaller than the Hauser-Feshbach component,
even at the highest neutron energies, and so we expect the impact on $P_{xn}$
to be small. For the $N>83$ nuclei, however, the pre-equilibrium spectrum is
similar in magnitude to the Hauser-Feshbach spectrum, and we expect that any
differences in the shapes of these spectra will impact $P_{xn}$.

The $P_{xn}$ values computed for these nuclei are given in \cref{tab:PxnSn132},
where calculations are performed assuming an internal transition hindrance
parameter $\delta=0.1$. It should be noted that these predictions are sensitive
to the $I_\text{GT}$ input, and their accuracy is further limited by
the neglect of first-forbidden transitions. In this work, we concern ourselves
less with the absolute $P_{xn}$ values, and more with the changes in $P_{xn}$
induced by the inclusion of a pre-equilibrium component in the primary neutron
spectrum.
The impact is illustrated in \cref{fig:DPxnSn132}, where the quantity
$\Delta P_{xn}\equiv P_{xn}^{\text{HF+PE}}-P_{xn}^{\text{HF}}$ is the
difference between the $xn$-emission probability predicted by the model that
includes pre-equilibrium decay and the Hauser-Feshbach model alone. As seen in
the figure, the greatest impact is on the more-exotic nuclei $^{134,135}$In, 
$^{133,134}$Cd, and $^{132,133}$Ag, with $P_{1n}$ decreasing by $\lesssim3\%$,
and $P_{2n}$ increasing to compensate, relative to the predictions of the pure
Hauser-Feshbach model. The impact is greater for a stronger hindrance parameter
$\delta$: increasing the likelihood of particle emission
relative to internal transitions in the early stages of the exciton model
evolution will increase the magnitude of the pre-equilibrium spectrum, and,
therefore, its relative contribution to the total spectrum.
By increasing $P_{2n}$ and decreasing $P_{1n}$, the multiplicity of neutrons
following $\beta$ decay of these nuclei increases by $\lesssim1.5\%$ relative
to the pure Hauser-Feshbach prediction. The relative difference in multiplicity
is shown in the inset of \cref{fig:DPxnSn132}.

\begin{figure}
\includegraphics[width=0.45\textwidth]{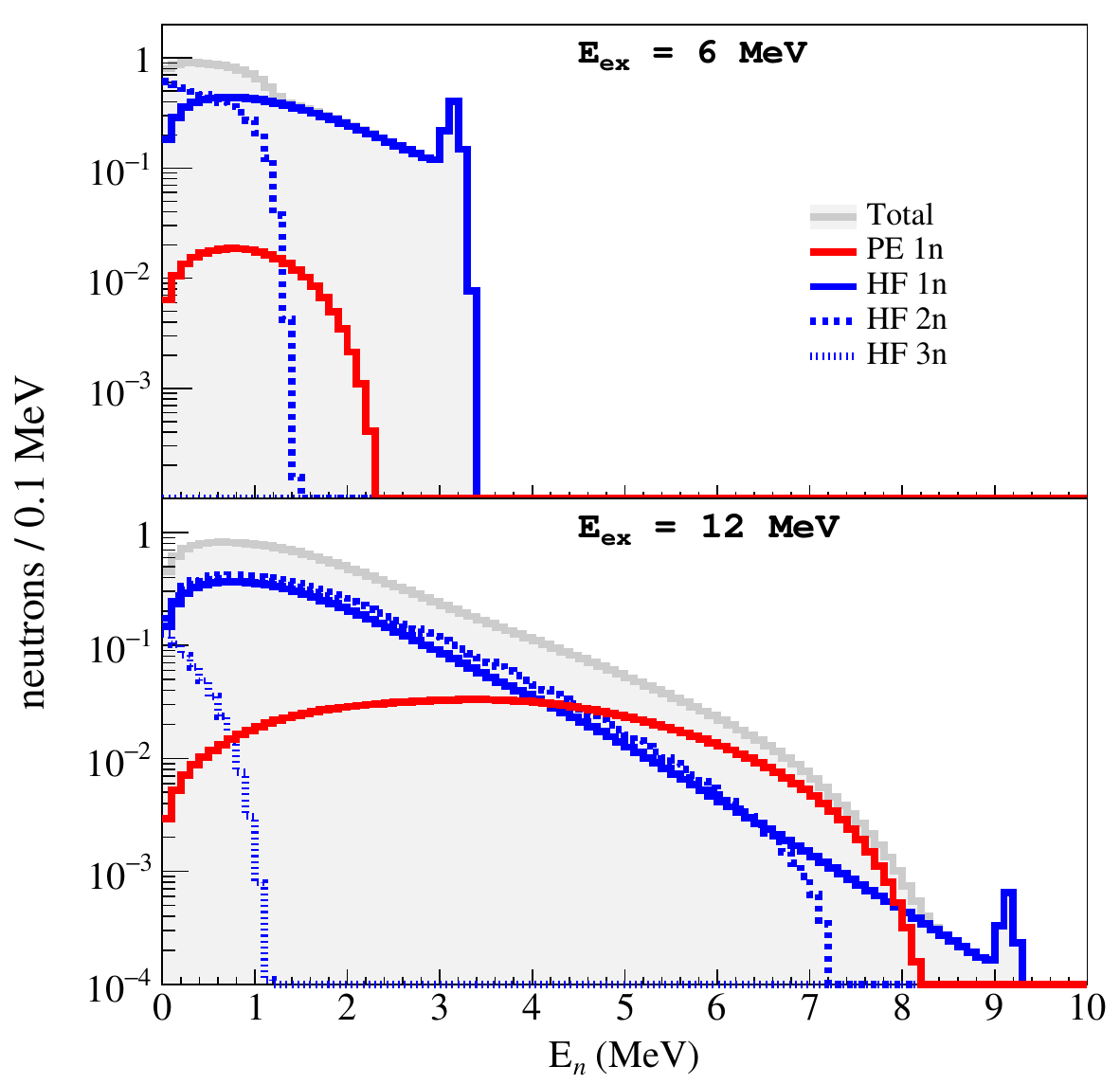}
\caption{
Neutron spectra for statistical and pre-equilibrium decay of states in
$^{132}$Cd at excitation energies of 6 MeV (top) and 12 MeV (bottom). The
pre-equilibrium spectra are calculated with an internal transition hindrance
factor $\delta=0.1$.
}
\label{fig:specHiLo}
\end{figure}

\begin{figure}
\includegraphics[width=0.31\textwidth]{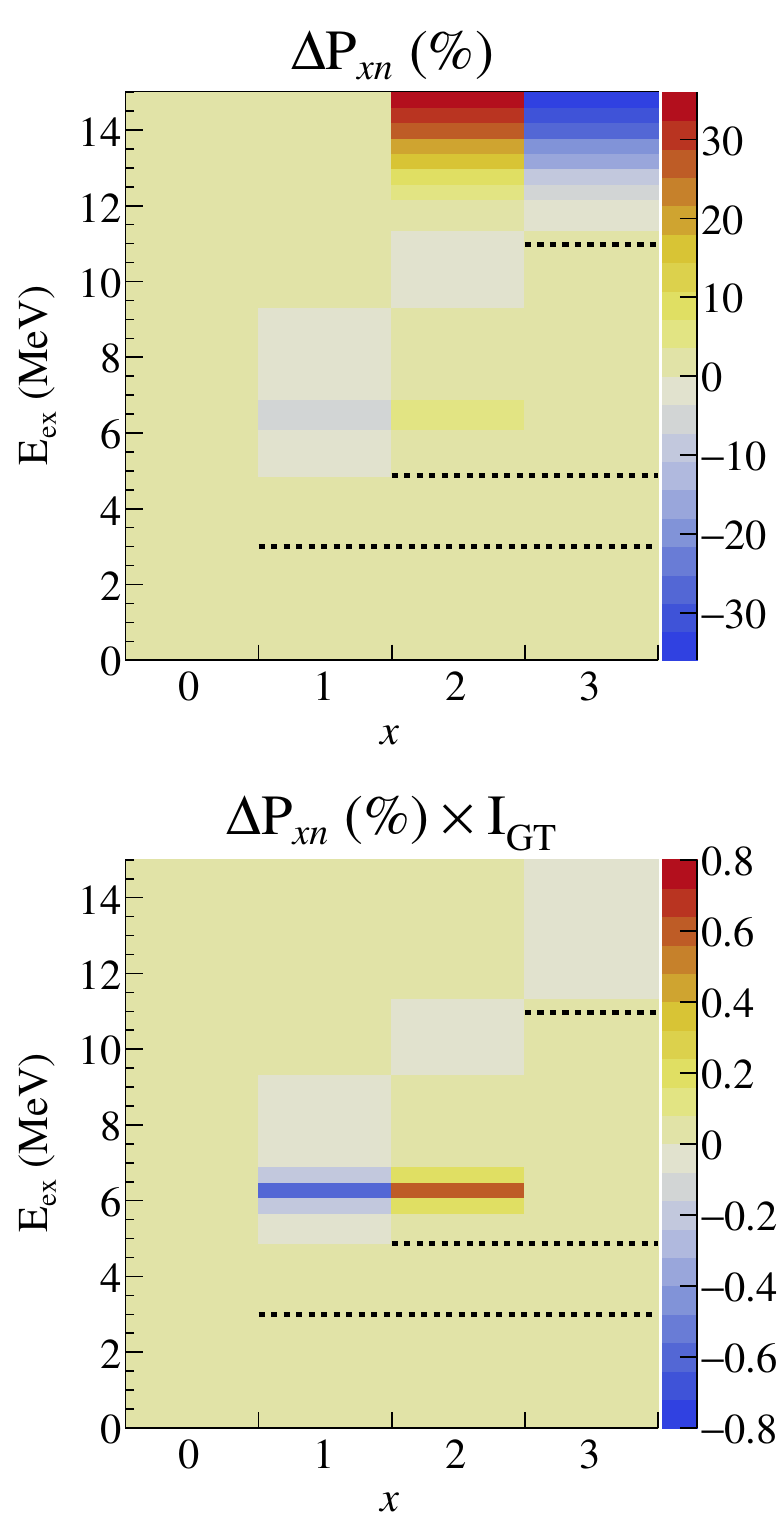}
\caption{
(Top) Difference in predicted $P_{xn}$ values for $^{132}$Ag with and without a
pre-equilibrium component, as a function of excitation energy in the
daughter $^{132}$Cd. Dashed lines are drawn at the $xn$-separation energies.
The calculations assume an internal transition hindrance parameter
$\delta=0.1$. (Bottom) Same as top, with $\Delta P_{xn}$ weighted by the
Gamow-Teller intensity $I_\text{GT}$ at each excitation energy.
}
\label{fig:DPxnExAg132}
\end{figure}

\begin{table}
\renewcommand{\arraystretch}{1.2}
\caption{$P_{xn}$ values (in \%) for nuclei in the region southeast of
$^{132}$Sn, computed using the exciton model for pre-equilibrium emission
together with the Hauser-Feshbach model for statistical emission. The
calculations assume an internal transition hindrance factor $\delta=0.1$. The
precursor $^{133}$Ag has a negligible $P_{4n} < 10^{-4}~\%$.}
\begin{tabularx}{0.45\textwidth}{@{\extracolsep{\fill}}ccccc}
\hline
& $P_{0n}$ & $P_{1n}$ & $P_{2n}$ & $P_{3n}$\\
\hline\hline
$^{132}$In &  60.2       &  39.8 &  $<10^{-4}$  &  $-$        \\
$^{133}$In &  $<10^{-1}$ &  99.5 &   0.5        &  $-$        \\
$^{134}$In &  $<10^{-3}$ &  75.8 &  24.2        &  $<10^{-5}$ \\
$^{135}$In &  $<10^{-3}$ &  75.3 &  24.2        &  0.5        \\
$^{131}$Cd &  77.7       &  22.3 &  $<10^{-2}$  &  $-$        \\
$^{132}$Cd &  $<10^{-3}$ &  99.6 &   0.4        &  $-$        \\
$^{133}$Cd &  $<10^{-2}$ &  83.0 &  17.0        &  $<10^{-3}$ \\
$^{134}$Cd &  $<10^{-1}$ &  88.1 &  11.4        &  0.5        \\
$^{130}$Ag &  16.5       &  82.6 &   0.9        &  $-$        \\
$^{131}$Ag &  $<10^{-1}$ &  95.4 &   4.6        &  $<10^{-3}$ \\
$^{132}$Ag &  $<10^{-4}$ &  34.2 &  65.4        &  0.4        \\
$^{133}$Ag &  $<10^{-3}$ &  50.9 &  43.3        &  5.8        \\
\hline
\end{tabularx}
\label{tab:PxnSn132}
\end{table}

We understand the cause of the change in $P_{xn}$ as follows, using
the case of $^{132}$Ag $\beta$-$xn$ decay as an illustrative example
(conclusions for the other nuclei in this region hold similarly). For neutron
emission from states at excitation energies below $B_{2n}$, the pre-equilibrium
spectrum has minimal impact on the $P_{0,1n}$ values, firstly, because the
magnitude of the pre-equilibrium spectrum is small relative to the
Hauser-Feshbach spectrum, and, secondly, because the branching ratio for
$\gamma$ emission from unbound states is negligible, so the dominant
determining factor for the $P_{0,1n}$ values is the fraction of $I_{\text{GT}}$
that lies above $B_{1n}$.

At higher $^{132}$Cd excitation energies, where multi-neutron emission channels
are open, pre-equilibrium emission that results in a change in the
shape of the primary neutron spectrum can have significant impact on the
subsequent neutron emissions. At energies between $B_{2n}$ and $B_{3n}$, the
pre-equilibrium spectrum is softer than the Hauser-Feshbach spectrum, as can be
seen in the top panel of \cref{fig:specHiLo}, and the lower average neutron
energy means that, on avarge, higher-energy states are populated in the
$\beta1n$ daughter $^{131}$Cd; there will be a greater fraction of the 
population above this daughter's neutron separation energy, leading to more
$2n$ emission relative to the Hauser-Feshbach prediction. This can be seen as
well in \cref{fig:DPxnExAg132}, where, at 6 MeV excitation energy in
$^{132}$Cd, we see the same decrease in $P_{1n}$ and corresponding increase in
$P_{2n}$.

At higher energies still, the pre-equilibrium spectrum becomes harder than the
Hauser-Feshbach spectrum, as shown in the bottom panel of \cref{fig:specHiLo}.
We therefore see the opposite effect: including a pre-equilibrium component
favors neutron decays of lower multiplicity. This can also be seen in
\cref{fig:DPxnExAg132}, where, above $B_{3n}$, we see $P_{2n}$ increase while
$P_{3n}$ decreases. The impact of pre-equilibrium emission is stronger at these
higher energies, since the pre-equilibrium and Hauser-Feshbach spectra have
similar magnitudes. When convoluted with $I_\text{GT}$, as in the bottom panel
of \cref{fig:DPxnExAg132}, the location of the Gamow-Teller peak at 6 MeV, just
above $B_{2n}$, means that the dominant competition is between $1n$ and $2n$
emission, with the softer primary neutron spectrum favoring $2n$ emission
more than in the pure Hauser-Feshbach model. Thus, we see an overall increase
in the $\beta$-delayed neutron multiplicity for $^{132}$Ag (and other nuclei
in this region) when including pre-equilibrium decays. However, since the
pre-equilibrium spectrum at the 6 MeV Gamow-Teller peak is roughly an order of
magnitude smaller than the Hauser-Feshbach spectrum, we expect an impact of
only a few percent on the $P_{xn}$ values; indeed, we see
$|\Delta P_{xn}| \lesssim 3\%$, as shown in \cref{fig:DPxnSn132}.

In the $^{132}$Sn region, where $\beta$-$xn$ data exist, the predicted changes
in $P_{xn}$ are comparable to the experimental uncertainties and smaller than
the sensitivity to the $I_\text{GT}$ input. A more stringent test of the model
would be to directly compare $\beta$-delayed neutron energy spectra and
excitation-energy-resolved decay paths reconstructed from neutron-$\gamma$
coincidences---especially when combined with $\beta$-feeding distributions
measured independently---rather than integrated observables like $P_{xn}$.
Energy correlations between the two neutrons in
$\beta$-$2n$ decay also offer a direct test, since the model presented in this
work involves a pre-equilibrium first neutron followed by a statistical
second neutron. Such measurements remain limited by the low two-neutron
coincidence efficiency of current detector arrays.

% --------------- %

\subsection{Proton-Deficient $N=125,126$ Isotones}

\begin{figure}
\includegraphics[width=0.48\textwidth]{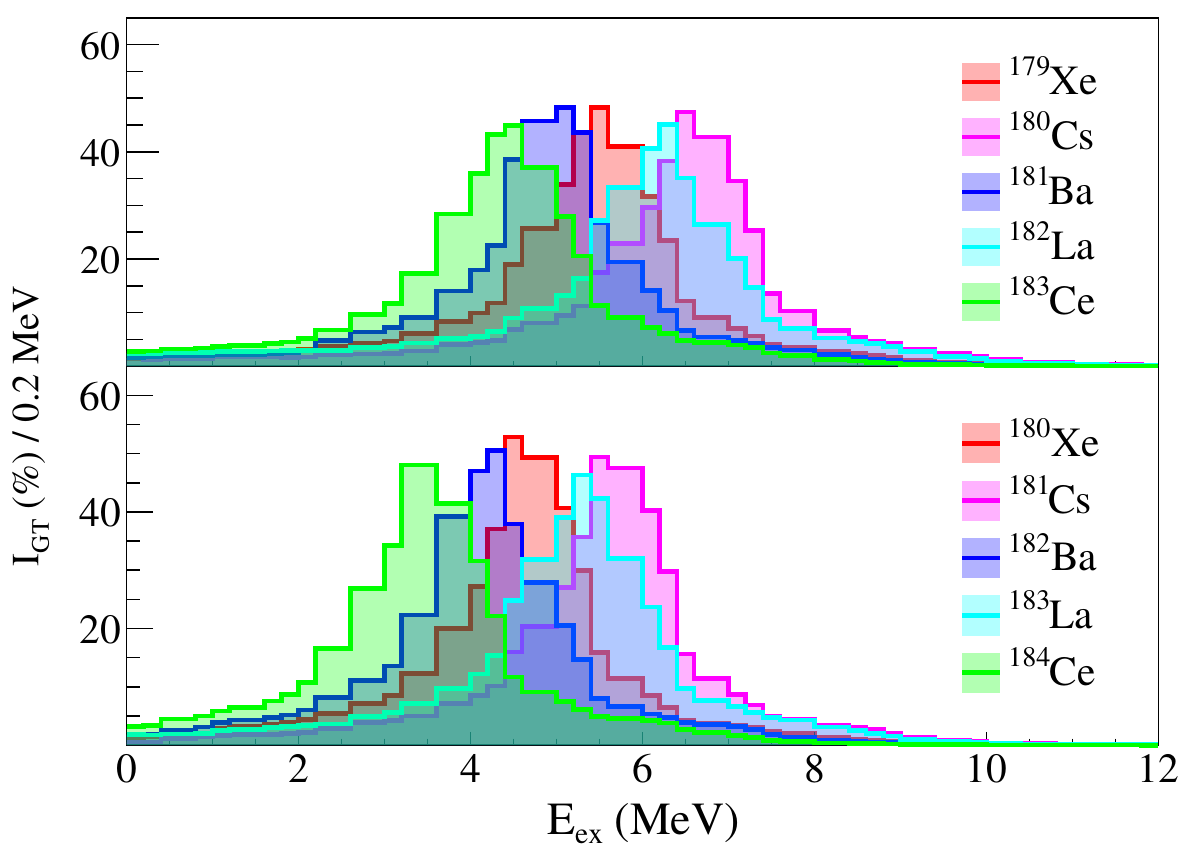}
\caption{
Gamow-Teller intensity distribution $I_\text{GT}$ for $\beta$ decay of the
indicated proton-deficient $N=125$ (top), 126 (bottom) isotones. The intensity
distribution is derived from the LSSM $B(\text{GT})$ reported in
\cite{Kum2024}.
}
\label{fig:IbetaN126}
\end{figure}

\begin{figure}
\includegraphics[width=0.5\textwidth]{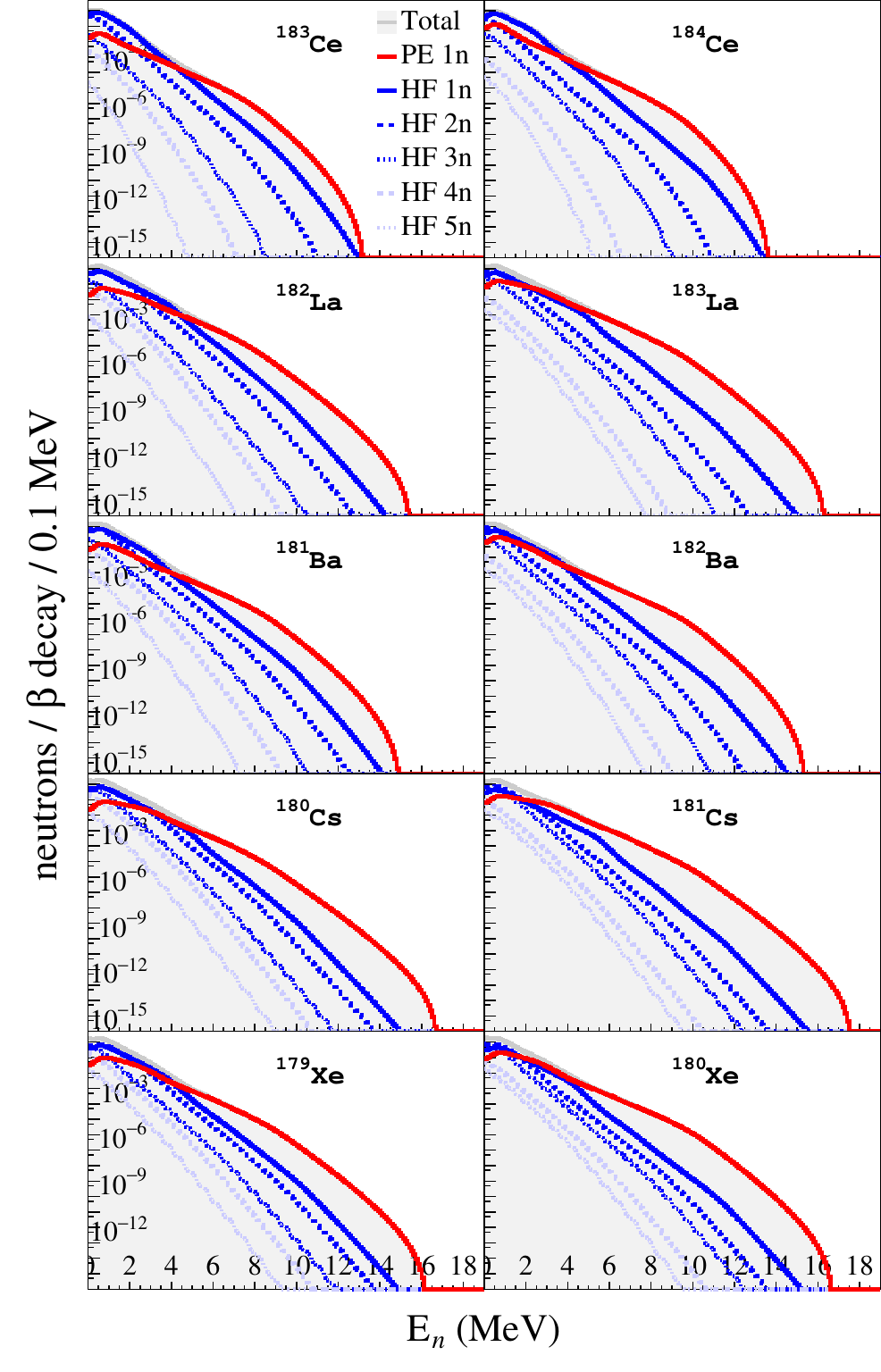}
\caption{
Pre-equilibrium (PE) and Hauser-Feshbach (HF) $\beta$-delayed neutron spectra
for proton-deficient $N=125,126$ precursors. See \cref{fig:specSn132} for
details. The spectra for neutron multiplicities higher than 5 are not shown,
due to their negligible branching ratios.
}
\label{fig:specN126}
\end{figure}

\begin{figure*}[t]
\includegraphics[width=1.0\textwidth]{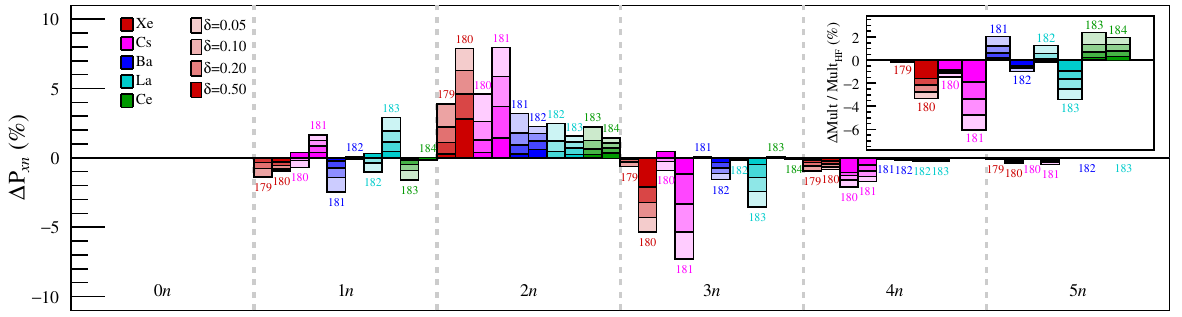}
\caption{
Difference $\Delta P_{xn}$ in predicted neutron-emission probabilities,
and relative difference in predicted neutron multiplicity, for calculations
with and without a pre-equilibrium component, for proton-deficient $N=125$, 126
precursors. See \cref{fig:DPxnSn132} for details.
}
\label{fig:DPxnN126}
\end{figure*}

\begin{figure}
\includegraphics[width=0.31\textwidth]{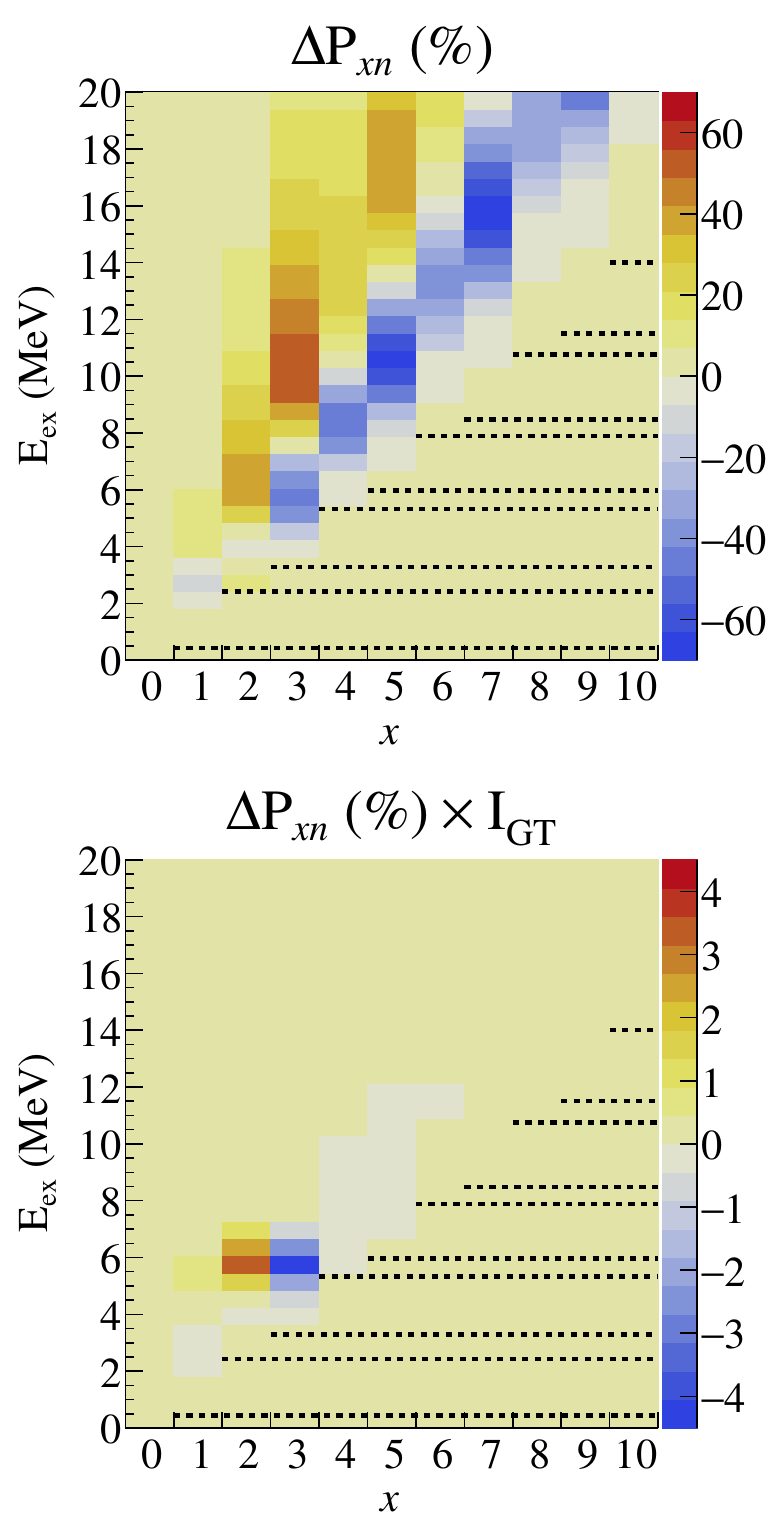}
\caption{
Difference in predicted $P_{xn}$ values for $^{181}$Cs with and without a
pre-equilibrium component (top), and the same, weighted by $I_\text{GT}$
(bottom). See \cref{fig:DPxnExAg132} for details.
}
\label{fig:DPxnExCs181}
\end{figure}

\begin{table}
\renewcommand{\arraystretch}{1.2}
\caption{$P_{xn}$ values (in \%) for proton-deficient $N=125$, 126 isotones,
computed using the exciton model for pre-equilibrium decay together with the
Hauser-Feshbach model for statistical decay. The calculations assume an
internal transition hindrance factor $\delta=0.1$. The probabilities for any
channels open beyond $5n$ emission are negligible, except for:
$P_{6n}(^{179}\text{Xe}) = 0.2 \%$ and
$P_{6n}(^{180}\text{Cs}) = 0.2 \%$.}
\begin{tabularx}{0.45\textwidth}{@{\extracolsep{\fill}}ccccccc}
\hline
& $P_{0n}$ & $P_{1n}$ & $P_{2n}$ & $P_{3n}$
	& $P_{4n}$ & $P_{5n}$\\
\hline\hline
$^{179}$Xe &  2.3 &  4.7 & 55.1 & 27.1 & 10.0 &  0.6 \\
$^{180}$Xe &  0.3 & 11.2 & 32.9 & 52.1 &  2.6 &  0.9 \\
$^{180}$Cs &  2.5 &  4.3 & 48.7 & 30.8 & 12.9 &  0.6 \\
$^{181}$Cs &  0.1 &  9.2 & 32.9 & 52.8 &  3.9 &  1.1 \\
$^{181}$Ba &  3.5 & 19.8 & 68.1 &  6.9 &  1.6 &  0.1 \\
$^{182}$Ba &  0.9 & 30.9 & 57.7 &  9.8 &  0.6 &  0.1 \\
$^{182}$La &  5.7 & 12.0 & 70.8 &  9.1 &  2.4 & <0.1 \\
$^{183}$La &  1.0 & 21.0 & 58.6 & 18.7 &  0.6 &  0.1 \\
$^{183}$Ce &  8.5 & 58.4 & 31.8 &  1.2 &  0.1 & <0.1 \\
$^{184}$Ce &  4.3 & 73.7 & 20.2 &  1.8 & <0.1 & <0.1 \\
\hline
\end{tabularx}
\label{tab:PxnN126}
\end{table}

We investigate the $\beta$-delayed multi-neutron emission probabilities for
proton-deficient $N=125,126$ isotones with $54\leq Z\leq 58$: nuclei for which
pre-equilibrium emission is expected to have a large impact.
For the $I_\text{GT}$ distributions in this region, we adapt
the $B(\text{GT})$ reported in \cite{Kum2024} from LSSM calculations using the
KHHE interaction \cite{Yua2022,War1991}. The $I_\text{GT}$ distributions
used in the present work are shown in \cref{fig:IbetaN126}. In this region,
no discrete states in the daughter nuclei are known.
We choose to investigate the most proton-deficient $N=125,126$ isotones covered
by the LSSM
calculations of \cite{Kum2024} because their large $Q_\beta \approx$ 20 MeV
 and low $B_{xn}$ \cite{Mol2016} maximize the phase space for multi-neutron
emission, making these nuclei extreme test cases for pre-equilibrium effects.
The $r$-process path crosses the shell closure at $N=126$ over a wide range of
proton numbers, from $Z\approx75$ down to $Z\approx55$ for the most
neutron-rich conditions, where the path approaches the dripline \cite{Arn2007}.
The $54\leq Z\leq58$ isotones studied here therefore represent the extreme,
drip-line end of the path, whose location carries a large systematic
uncertainty \cite{Erl2012,Neu2020}.

The pre-equilibrium and Hauser-Feshbach $\beta$-delayed neutron spectra are
shown in \cref{fig:specN126} for calculations with an internal transition
hindrance parameter $\delta=0.1$. In all cases, the primary neutron spectrum is
dominanted by the pre-equilibrium component at neutron energies above
$\approx 2$ MeV, and the total neutron spectrum is made moderately harder by
including pre-equilibrium emission. The $P_{xn}$ values derived from these
calculations are given in \cref{tab:PxnN126}. For Xe and Cs isotones, the
strongest competition is between $2n$ and $3n$ emission, while for the
less-proton-deficient Ba, La, and Ce isotones, the competition is between $1n$
and $2n$ emission.

\Cref{fig:DPxnN126} shows the impact of including pre-equilibrium emission
on the predicted $P_{xn}$ values, similar to \cref{fig:DPxnSn132}.
The largest impact $|\Delta P_{xn}|\lesssim8\%$ is seen for
the most proton-deficient isotones, Xe and Cs, with an increase in $P_{2n}$ and
corresponding decrease in $P_{3n}$. Thus, the overall $\beta$-delayed neutron
multiplicity decreases by $\lesssim6\%$ relative to the Hauser-Feshbach
prediction, as shown in the inset of \cref{fig:DPxnN126}. This is opposite to
the trend seen in the $^{132}$Sn region, where a (small) relative
\textit{increase} in the neutron multiplicity was observed. The different
behavior seen in the two regions is due to the location of the Gamow-Teller
peak relative to the various neutron binding energies: for the $N=125$, 126 Xe
and Cs isotones, the Gamow-Teller peak lies well above $B_{3n}$ (see
\cref{fig:IbetaN126,fig:DPxnExCs181}), in a region where the
pre-equilibrium spectrum is harder than the Hauser-Feshbach spectrum, so the
dominant effect of pre-equilibrium emission is to increase the relative
population of states below $B_{3n}$ (but above $B_{2n}$), compared to the
Hauser-Feshbach model. In contrast, in the $^{132}$Sn region, the Gamow-Teller
peak lies just above $B_{2n}$, where the pre-equilibrium spectrum is softer
than the Hauser-Feshbach spectrum, and the neutron multiplicity tends to
increase when including a pre-equilibrium component. These effects can be
easily seen in \cref{fig:DPxnExCs181}, where $\Delta P_{xn}$ is shown as a
function of excitation energy in $^{181}$Ba, populated in the $\beta^-$ decay
of $^{181}$Cs. The bottom panel of the figure shows $\Delta P_{xn}$ convoluted
with the Gamow-Teller intensity $I_\text{GT}$. Clearly, at the Gamow-Teller
peak $\approx 6$ MeV, the pre-equilibrium decay has the largest impact on $2n$
and $3n$ emission. At higher energies, the impact shifts to higher neutron
multiplicities.

%=============================================================================%

%-------------------------%
\section{Conclusions\label{sec:conclusions}}
%-------------------------%

In this work, we have described the implementation in \textsc{BeoH} of a
two-component exciton model for pre-equilibrium neutron emission from nuclear
states populated in $\beta^-$ decay. Primary- and multi-neutron emission after
equilibration of the daughter nucleus was computed using the existing
Hauser-Feshbach code in \textsc{BeoH}. We investigated the impact of including
a pre-equilibrium component in the primary neutron emission spectrum on the
resultant neutron emission probabilities $P_{xn}$. In the region southeast
of $^{132}$Sn, we observed a decrease in $P_{1n}$ of $\lesssim 3\%$ and a
corresponding increase in $P_{2n}$ relative to the predictions of the pure
Hauser-Feshbach model, resulting in a small $\lesssim 1.5\%$ relative increase
in the overall $\beta$-delayed neutron multiplicity. In the proton-deficient
$N=125$, 126 isotones, we observed an increase in $P_{2n}$ of $\lesssim 8\%$
and a corresponding decrease in $P_{3n}$ relative to the Hauser-Feshbach
predictions, resulting in a moderate $\lesssim 6\%$ relative decrease in the
multiplicity. These results are sensitive to the hindrance of the internal
transition rates over particle emission rates in the early stages of the
exciton model evolution; more-hindered internal transition rates result in a
stronger pre-equilibrium component of the total spectrum, and therefore, in a
larger impact.

The effects of pre-equilibrium decay on the $\beta$-delayed neutron
emission probabilities and neutron multiplicities in the two regions of the
nuclear chart considered here could impact the results of models for
$r$-process nucleosynthesis and $\beta$-$xn$ decay of actinide fission
fragments. A dedicated sensitivity study would be needed to assess the impact
of pre-equilibrium neutron emission on other observables of interest, and in
other regions of the nuclear chart. Future improvements to the described
exciton model for $\beta$-$xn$ decay might include the treatment of
pre-equilibrium multi-neutron emission, or of neutron emission from states
populated in first-forbidden $\beta^-$ decay.
Comparisons of the theoretical model presented in this work with experimental
neutron spectra will become possible as more exotic, neutron-rich isotopes are
produced at accelerator facilities. In cases where the
measured $P_{xn}$ values differ from the predictions of the Hauser-Feshbach
model, it may be prudent in these exotic regions to consider pre-equilibrium
neutron emission as a possible solution to such discrepancies.
We conclude by noting that, in
the over half-century since the introduction of the Hauser-Feshbach model in
1952, the theoretical description of neutron emission following $\beta^-$ decay
has received little development outside the fundamental framework of
statistical decay of compound-nucleus states. We hope that further
investigation into non-statistical models for $\beta$-$xn$ decay, such as the
exciton model described here, will provide valuable insights into a decay mode
that is critical to many scientific and industrial applications.
\newline

% ----------- %

This work was carried out under the auspices of the National Nuclear Security
Administration of the U.S.\ Department of Energy at Los Alamos National
Laboratory under Contract No.\ 89233218CNA000001.
N.B.\ acknowledges Laboratory Directed Research and Development Program support
for Project No.\ 20261415CR SEA-STU: Seaborg Institute: Center for Advancing
Actinide Science and Technology at Los Alamos National Laboratory.

%-------------------------%
% backmatter
\bibliography{ref}

\end{document}